# Probing the Faintest Stars in a Globular Star Cluster

Harvey B. Richer[1#], Jay Anderson[2], James Brewer[1], Saul Davis[1], Gregory G. Fahlman[3], Brad M.S. Hansen[4], Jarrod Hurley[5], Jasonjot S. Kalirai[6], Ivan R. King[7], David Reitzel[4], R. Michael Rich[4], Michael M. Shara[8], Peter B. Stetson[3]

[1]*Department of Physics and Astronomy, University of British Columbia, Vancouver, BC, Canada*

[2]*Department of Physics and Astronomy, Rice University, Houston, Texas*

[3]*National Research Council, Herzberg Institute of Astrophysics, Victoria, BC, Canada*

[4]*Division of Astronomy and Astrophysics, UCLA, Los Angeles, CA*

[5]*Department of Mathematics and Statistics, Monash University, Clayton, Australia*

[6]*Department of Astronomy and Astrophysics, UCSC, Santa Cruz, CA, Hubble Fellow*

[7]*Department of Astronomy, University of Washington, Seattle, Washington*

[8]*Department of Astrophysics, American Museum of Natural History, New York, NY*

[#]*To whom correspondence should be addressed. E-mail: richer@astro.ubc.ca*

**NGC 6397 is the second closest globular star cluster to the Sun. Using 5 days of time on the Hubble Space Telescope, we have constructed the deepest ever color-magnitude diagram for this cluster. We see a clear truncation in each of its two major stellar sequences. Faint red main sequence stars run out well above our observational limit and near to the theoretical prediction for the lowest mass stars capable of stable hydrogen-burning in their cores. We also see a truncation in the number counts of faint blue stars, namely white dwarfs. This reflects the limit to which the bulk of the white dwarfs can cool over the lifetime of the cluster. There is also a turn towards bluer colors in the least luminous of these objects. This was predicted for the very coolest white dwarfs with hydrogen-rich atmospheres as the formation of $H_2$ causes their atmospheres to become largely opaque to infrared radiation due to collision-induced absorption.**

When stars are born, they contract under gravity, increasing their central temperatures and densities. If a star is massive enough, at least 0.08 that of the Sun (80 times the mass of Jupiter), the center will eventually get hot enough to allow self-sustaining nuclear energy generation at a level sufficient to halt the contraction. For lower mass objects, the contraction is halted instead by electron degeneracy, the quantum-mechanical repulsion between electrons in dense media. For these "brown dwarfs", the rate of energy generation never gets large enough to be self-sustaining and they fade away within a billion years (*1*). On the other hand, for low mass stars just above the brown dwarf mass limit, the balance between gravity and thermal equilibrium is long-lived, and these stars are able to shine for much longer than the current 13.7 billion year age of the Universe (*2*). The sharp division between transient and effectively infinite lifetime populations is termed the hydrogen-burning mass limit and is a cornerstone of stellar evolutionary theory. However, due to the faintness of the stars just above the limit, this edge has not previously been solidly identified in any old stellar population such as globular clusters or halo field stars.

At the other end of the stellar mass spectrum, stars with masses ranging from about 1 through 7 solar masses consume their nuclear fuel on a timescale shorter than the age of the Universe, burning their original hydrogen to carbon and oxygen, ejecting their outer layers and becoming white dwarfs. These stellar remnants are also supported by electron degeneracy pressure and slowly cool over time as they radiate their reservoir of thermal energy left over from the previous stages of nuclear burning. The accompanying slow fading, at approximately constant radius, traces out the so-called white dwarf cooling sequence in the color-magnitude diagram (CMD). At faint magnitudes, this sequence is expected to end, with the magnitude of the cut-off indicating the luminosity to which the oldest white dwarfs have faded. With the aid of theoretical models, this luminosity can then be used to derive the age of the cluster (*3,4*).

In recent years, it has been appreciated that the white dwarf cooling sequence is expected to turn back to the blue at faint magnitudes (*5-9*). As white dwarfs with hydrogen-rich atmospheres cool below temperatures of 4000 K, they exhibit the collision-induced absorption (CIA) of molecular hydrogen (*10*). CIA is strongest at near-infrared wavelengths, which suppresses the flux near 1 micron (*11*) and causes the optical colors to become bluer as the star cools, rather than redder as might be expected from a blackbody. CIA arises because hydrogen molecules can form in these cool atmospheres. Being symmetric, $H_2$ does not have a dipole moment and will only very weakly absorb radiation via quadrupole transitions (*10*). A dipole moment may be induced, however, during an $H_2$–$H_2$, $H_2$–H, or $H_2$–He collision (*12*), leading to CIA. The stars continue to get bolometrically fainter and cooler as they evolve on the blueward track in the CMD known as the blue hook. The spectral signature of CIA has been seen in a few individual field white dwarfs (*13-15*) but never as a feature in the cooling sequence of a stellar cluster. The expected cluster CIA signature is a blue hook in the cooling sequence. We note that this is a temperature effect, seen at faint magnitudes, and distinct from the feature observed in the open cluster NGC 6791 (*16*). The latter feature occurs in younger populations and is a consequence of the smaller radii and slower cooling of more massive white dwarfs at early times, as shown in (*17, Fig 17*).

In an attempt to locate the limits of both the faint main sequence star population and the faint white dwarf population, we obtained the deepest ever globular star cluster images with the Hubble Space Telescope in NGC 6397.

NGC 6397 is located in the southern constellation Ara. It is seen projected in the direction of the Galactic Center, is located at a distance of about 8500 light years from the Earth (18), and its stars have a heavy element content that is ~1% that of the Sun (19).  It was discovered by Abbe Nicholas Louis de la Caille during his 2 year sojourn at the Cape of Good Hope in 1751-1752. In his original catalogue it bears the name Lacaille III.11. It was the twelfth such cluster discovered.  Over 250 years later, we imaged a single field in NGC 6397 with the Advanced Camera for Surveys (ACS) on the Hubble Space Telescope (HST) for a total of 126 orbits (4.7 days) in order to characterize the faintest globular cluster stellar populations. Our observations were carried out 5′ SE of the cluster core. They overlap pre-existing Wide Field Planetary Camera 2 (WFPC2) images that we used to select by common proper motion (angular motion in the plane of the sky), a clean sample of cluster stars largely devoid of Galactic foreground and background stars. The exposures were taken through two filters, F606W and F814W, where the number indicates the central wavelength of the filter in nanometers.

The data set consists of 252 exposures totaling 179.7 ksec in  F814W and 126 exposures of total exposure 93.4 ksec in F606W. This 2/1 ratio was chosen as the primary science goal of the program was exploration for the coolest white dwarfs. If cool white dwarfs do indeed turn blue at faint magnitudes *(5-9)*, they will be very faint on the longer wavelength images thus requiring increased exposure time to detect them. To find the faintest possible sources, we scoured each image for indications of where the faintest stars might be located.  The most we could hope for from the very faintest ones is that they would push their central pixels up above the noise in some significant number of images.  We studied the noise in the images via artificial star tests, and concluded that a detection in 90 out of 252 F814W images constituted a 3σ detection above the background. Figure 1 is an image of part of our field which was constructed by combining together the exposures in the two filters. The pair of  inserts provide expanded views of two stars at the extremities of  stellar evolution: the least luminous hydrogen-burning cluster star observed (bottom), and a white dwarf that is cool enough to lie in the blue hook part of the cooling sequence.

We constructed a CMD (Fig 2 (a)) from these data including all sources that: (1) generated local maxima in at least 90 out of 252 F814W images; (2) were the most significant sources within 7.5 ACS pixels, and (3) passed morphology tests indicating that they are point-like objects. In this diagram, brighter stars have small values of the F814W magnitude and red stars will have larger positive (F606W – F814W) colors. Some unresolved faint blue galaxies likely remain in the sample bluer than (F606W – F814W) = 1 and fainter than F814W = 27. We estimated the photometric errors for white dwarf stars using measurements of artificial stars of known magnitude inserted into the individual exposures and recovered by our finding and measuring algorithms. In addition, we considered the completeness for stars as a function of the F814W magnitude, that is, the fraction of stars that we recovered in the ACS images in artificial star tests. The completeness is based solely on the recovery fractions in the F814W data and hence applies equally well to both white dwarfs and main sequence stars. Completeness fractions in excess of 50% are normally considered acceptable.

Fig 2 (a) is dominated by a narrow continuous cluster sequence spanning at least 14 magnitudes in the F814W filter. The sequence extends from a sparse giant branch at the brightest F814W magnitudes, through the main sequence turnoff near F814W = 15.7, and down the main sequence to faint objects, with a possible termination in a cloud of field stars near F814W = 24. The turnoff region corresponds to stars that have recently completed hydrogen-burning in their cores. They will shortly evolve into red giants, a phase of evolution lasting about 100 million years, before they eject their atmospheres and become white dwarfs. The scatter around this narrow sequence is produced by stars not associated with the cluster. These are stars from the Galactic disk and bulge which are found along the line of sight to NGC 6397. This population is large due to the low Galactic latitude (-12 degrees) of NGC 6397.

The most novel feature of the diagram, however, is the white dwarf cooling sequence, which begins at about F814W = 22.5 and extends to F814W = 27.2, where it then 'hooks' toward the blue. The sequence appears to be largely truncated fainter than F814W = 27.8. This truncation reflects the limit to which the bulk of the hydrogen-rich white dwarfs can cool over the lifetime of the cluster. For this reason it will be a powerful diagnostic in determining the age of NGC 6397. However, this truncation does not necessarily mean that there are no white dwarfs in the cluster fainter than this limit. Over a long period of time, both massive white dwarfs (those with masses in excess of about 0.8 that of the Sun) and those with helium-rich atmospheres, cool more quickly and to fainter magnitudes than do white dwarfs of modest mass (0.5 – 0.6 solar masses) with hydrogen-rich atmospheres. As a consequence of electron degeneracy, more massive white dwarfs have smaller radii and hence larger densities. They will thus develop a crystallized core earlier than low mass white dwarfs causing them to enter a regime know as Debye cooling before the lower mass ones. In this phase of evolution, the star effectively loses its ability to store thermal energy and it cools very rapidly. Because helium does not form molecules (which would have a vibrational mode), white dwarfs with pure helium atmospheres do not suffer from CIA. Their atmospheres are thus less opaque to infrared radiation and they cool rapidly. Hence both massive and helium-rich white dwarfs may have cooled to below our observational limit.

We examined the details of the CMD further by analyzing the stars whose proper motions match that of the cluster (Fig 2(b) and Fig 3). Pre-existing WFPC2 images, which were used to proper motion clean the data, overlap only 60% of the area of the ACS field and are much shorter in exposure time (F814W exposure times of 3960, 7440 and 5200 sec in each of the 1994, 1997 and 2001 data sets compared to 179.7 ksec with the ACS in 2005) so that their utility (particularly for the very faintest stars) was limited. Distinguishing between cluster and field stars is a serious issue both for stars along the main sequence and stars in the white dwarf region. The main sequence stars become very red at faint magnitudes so they may be found on the F814W images but could be unmeasurable on the bluer F606W frames. For the white dwarfs, if they do indeed become bluer as they cool (*5-9*), we might expect to lose them on the F814W frames. Nevertheless, the major features are preserved and are much cleaner in Fig 2 (b) compared to 2 (a). The main sequence is now seen to continue to F814W ~ 26.0 at (F606W – F814W) ~ 4.0. An F814W magnitude of 26.0 and a F606W magnitude of 30.0 correspond to absolute magnitudes of 13.6 and 17.4 respectively. These absolute magnitudes are more than a full magnitude fainter than the faintest known metal-poor ('Population II') field stars (*20, 21*) suggesting that there remains a population of extremely dim metal-poor stars awaiting discovery in the halo of the Galaxy. The cluster main sequence has two remarkable features: one at F814W ~ 24 where the number counts of stars declines rapidly (noted by (*22*)), the second at F814W ~ 26 where the main sequence appears to terminate. This latter feature is not caused by incompleteness (Fig 3). Except for a few stars that scatter well away from the main sequence locus (all these are also outliers in the proper motion distribution so they are likely interlopers from the field population), there appears to be a complete lack of cluster main sequence stars fainter than F814W = 26. The presence of an extensive white dwarf population as faint as F814W = 27.8, as well as numerous faint field stars, shows that if a significant population of main sequence stars fainter than F814W = 26 were present, we would have found it.

We quantify this result by examining the proper motion distribution for faint red and blue stars (Fig 3). NGC 6397 is moving relative to the field stars. Over the course of 10 years since the first archival data were taken, the cluster stars have moved by almost 3 ACS pixels with respect to the bulk of those in the field. A displacement this large is trivial to measure provided we can find the star in the archival data. Figure 3 displays the proper motions for all objects with respect to the cluster as a function of F814W magnitude. Panel (a) illustrates the proper motion cuts we made to select cluster stars. These are all those to the left of the red line which is set by fitting a Gaussian function to the total proper motion distribution in bins 1 magnitude wide and selecting those stars that lie within two standard deviations from the mean. The remaining plots are proper motions along the detector axes for stars redder than (F606W − F814W) = 1.5 in the first three and for blue stars with (F606W − F814W) < 1.5 in the last one. The circle shown in each plot is the proper motion cut appropriate to that magnitude. In panel (b), the circle has a radius of 0.5 ACS pixels while it is 1.3 pixels for panels (d) and (e). By contrast, proper motion errors in x and y range from 0.05 pixels at F814W = 22 to 0.6 pixels at F814W = 26. It is clear that clumps corresponding to cluster stars are seen in the first two plots, but that no concentration is seen for red main sequence stars with F814W > 26. This coincides with the impression in Fig 2 (b) that there are no obvious cluster stars fainter than this limit. The final panel illustrates the displacements for stars with (F606W − F814W) < 1.5 and F814W > 26, which includes cluster white dwarfs, blue field stars, and unresolved blue galaxies. Clearly there is a sizeable component of stars moving with the cluster at these faint magnitudes (the white dwarfs), so that the deficiency seen for the main sequence stars is real and not due to incompleteness.

The apparent termination of our observed main sequence lies close to the predicted hydrogen-burning limit (Fig 2 (b) filled square); the lowest mass star capable of stable nuclear fusion of hydrogen in its core is 0.083 solar masses at the low heavy element abundance of NGC 6397 (*23*). We caution, however, that different models may yield different results. At masses just above the hydrogen-burning limit, the relation between the mass and luminosity of a star is very steep; an extremely small change in mass results in a large change in luminosity. This is because at low masses, electron degeneracy pressure becomes important in supporting the star. The temperature no longer increases with increasing pressure as it does for a classical gas. This has the effect of decreasing the central temperature and hence the nuclear energy generation rate (*24*). As stars approach the hydrogen-burning limit in mass, we thus expect to see a steep decline in their number counts as a function of luminosity. This makes it highly improbable that any stars will be found 'at' the hydrogen-burning limit.

The blue hook feature in the white dwarf cooling sequence remains after proper motion cleaning, leading us to conclude that we are indeed seeing the predicted CIA signature (*5-9*). We quantify the nature of the hook by deriving an empirical cooling sequence. We fit the observed color distribution in F814W magnitude bins of width 0.25 with a model of the dispersion in color determined from artificial star tests. This results in a relation between color and magnitude that is purely empirical and that can be compared to any theoretical models of the cooling. In Fig 4, we show the comparison between this relation and one particular theoretical cooling sequence. Note that there is indeed a blue hook in the empirical relation and there is excellent agreement between this and the model curve. This concordance suggests that further detailed modelling of these cool white dwarfs will provide a strong constraint on the white dwarf cooling age of NGC 6397.

In 1984 it was suggested to the European Space Agency by the late Vittorio Castellani and Vittoria Caloi that HST be used to locate the end of the hydrogen-burning main sequence in the nearest globular star clusters. Twenty-two years after that challenge to the astronomical community, we believe it has been answered. In addition, the blue hook and a truncation in the white dwarf cooling region have also been located. It now remains to exploit these features to learn more about the structure of low-mass stars, brown dwarfs, the original massive stellar component in globular clusters, white dwarf atmospheres and cooling, and ultimately, the ages of these oldest known stellar populations.

25. HBR thanks UCLA for support during his extended visit during which time most of this paper was written. He would also like to thank the US-Canada Fulbright Fellowship Committee for the award of a fellowship during his stay at UCLA. The research of HBR is largely funded by the Natural Sciences and Engineering Research Council of Canada, but support for this project was also provided by the University of British Columbia. JA received support from NASA/HST through grant GO-10424 as did BMSH, IRK, JSK, RMR and MMS. JSK is also supported by NASA through a Hubble Fellowship. HBR would like to thank I. Ozier for fruitful discussions on CIA.

26. This research is based on observations with the NASA/ESA Hubble Space Telescope, obtained at the Space Telescope Science Institute, which is operated by the Association of Universities for Research in Astronomy, Inc., under NASA contract NAS5-26555. These observations are associated with proposal GO-10424.

**Figure 1** A section of the observed field in NGC 6397 covering 94″ x 127″. This is 29% of the entire field of our observations. Directions on the sky are indicated as is the scale in arc seconds. The image is a composite of exposures from the HST/ACS wide field camera in F606W and F814W. The inserts are 10″ x 10″ and show detail for the faintest hydrogen-burning star observed in the cluster (upper), and a white dwarf along the blue hook part of the cooling sequence. Although the field is in a globular star cluster with many bright stars, the ACS has such a tight point spread function with relatively little scattered light, that it is still possible to see faint external galaxies right through the cluster. [See Published Article in Science for Full Resolution Image]

**Figure 2** The CMD of NGC 6397 without proper motion selection of cluster members (a), and after cleaning the data using proper motions ((b) and Fig 3) to exclude non-cluster stars in the Galactic disk and bulge from the sample. Most of the faint galaxies and artefacts have been removed from (a) by requiring that the found objects pass certain morphological tests. The 1σ photometric errors as a function of magnitude are indicated for white dwarfs by the error bars at F814W = 22.5, 24.5, 26.5 and 28.0, as well as the percentage of stars (either white dwarfs or main sequence) found in the ACS images at each of these magnitudes from artificial star tests. In (b) the cluster main sequence appears to terminate at F814W = 26, (F606W - F814W) = 4. The few scattered objects at lower luminosity are at the extremes of the proper motion selection criteria and are likely interlopers from the large field population. The two "x"s to the right indicate the F814W magnitudes of the stars that passed the proper motion cuts but whose F606W magnitudes were not measured. The filled square is the location of the hydrogen-burning limit for one particular set of theoretical models *(23)*. [See Published Article in Science for Full Resolution Image]

**Figure 3** Panel (a) – The F814W magnitude plotted against total stellar proper motion displacement (scaled to 10 years) with respect to the cluster. The units are ACS pixels that project to 0.05″ on the sky. The red line shows the selection criterion, all stars to the left of the line are considered to be cluster members. Panels (b) through (e) – Displacements in the x and y directions on the detector (-x is approximately north). The NGC 6397 stars are those in the tight clump at (0,0) while the bulk of the field stars are located in the diffuse clump about 3 pixels away. [See Published Article in Science for Full Resolution Image]

**Figure 4** The white dwarf region of the full CMD (i.e. not proper motion cleaned) overlaid with the empirical cooling sequence (red dots with 1σ error bars) derived as described in the text. We use the full CMD here in lieu of the proper motion cleaned one in order to avoid artificial truncation of the cooling sequence from losses of stars from the shorter exposure earlier epoch data. The solid blue curve is a theoretical cooling sequence using the atmospheric models of Bergeron et al. *(7)* and a cooling model for 0.5 solar mass white dwarfs of Hansen *(6)*. [See Published Article in Science for Full Resolution Image]

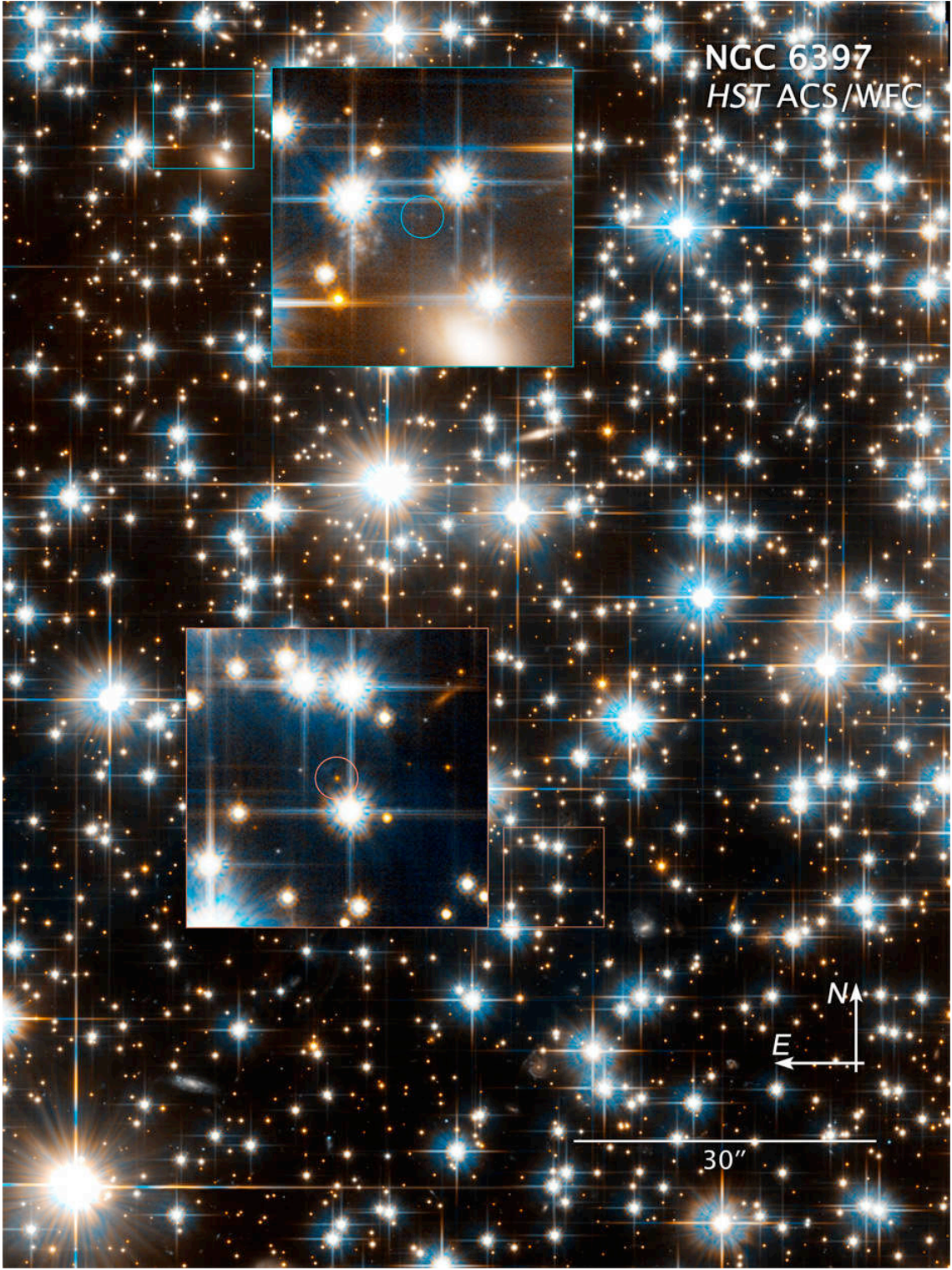

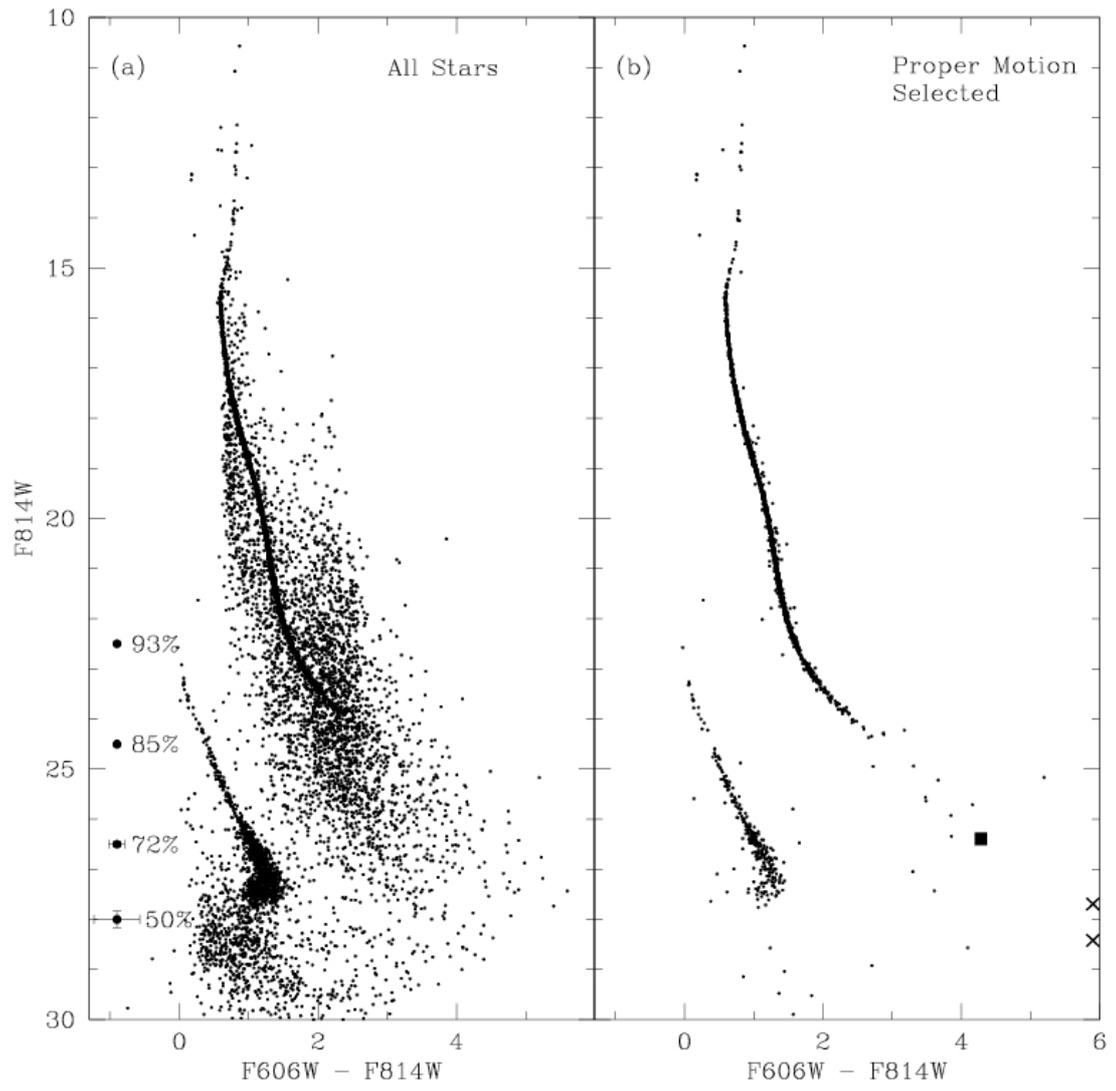

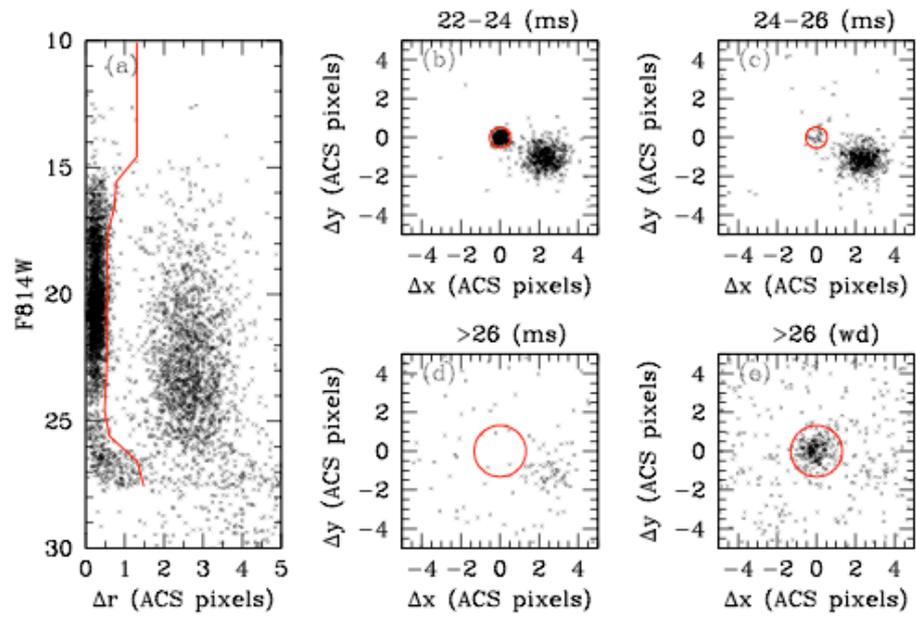

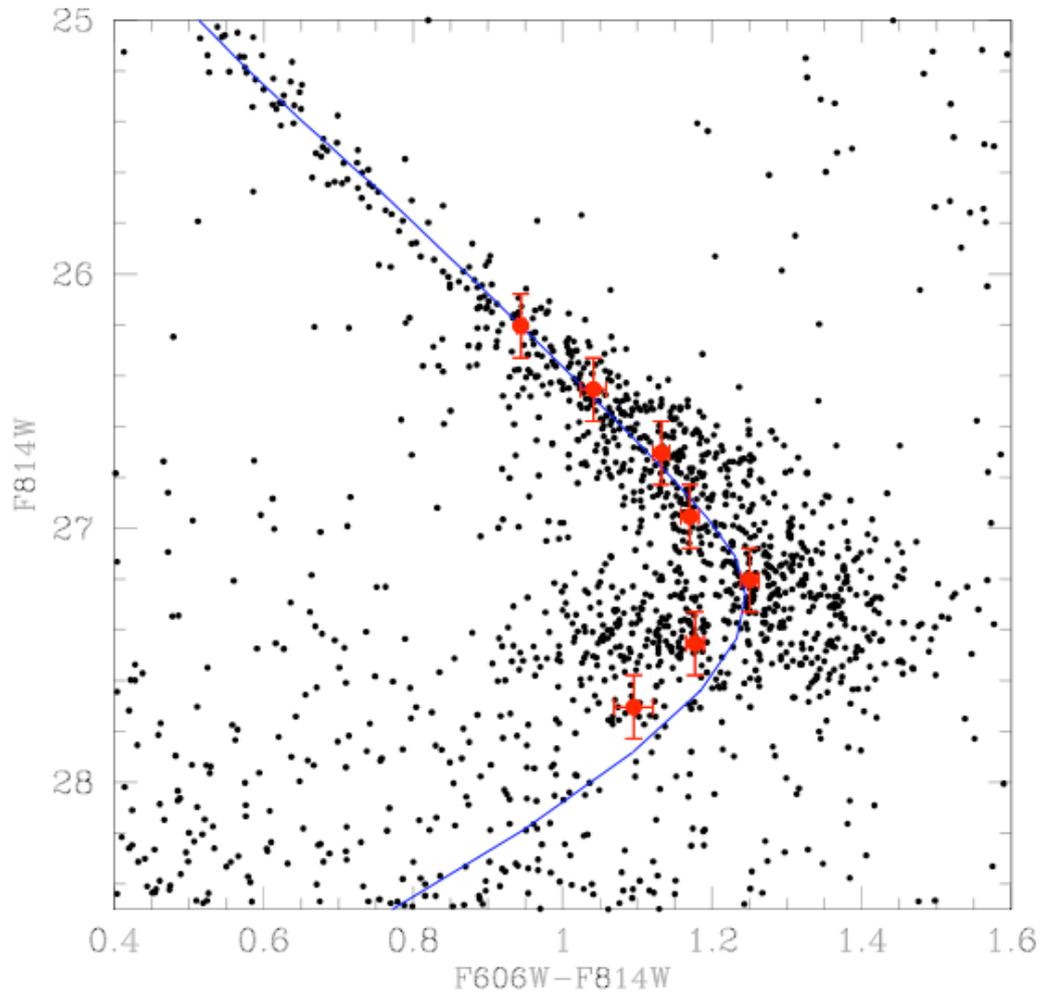